\documentclass[preprint,12pt,authoryear]{elsarticle}

\usepackage{amssymb}
\usepackage{amsmath}
\usepackage{booktabs}
\usepackage{url}

\journal{Computerized Medical Imaging and Graphics}

\begin{document}

\begin{frontmatter}



\title{WindowNet: Learnable Windows for Chest X-ray Classification}


\author[label1]{Alessandro Wollek\corref{alessandro.wollek@tum.de}}
\author[label2]{Sardi Hyska}
\author[label2]{Bastian Sabel}
\author[label2]{Michael Ingrisch}
\author[label1]{Tobias Lasser}

\affiliation[label1]{organization={Munich Institute of Biomedical Engineering, and School of Computation, Information, and Technology, Technical University of Munich},
            addressline={Boltzmannstr. 11},
            city={Garching},
            postcode={85748},
            state={Bavaria},
            country={Germany}}
\affiliation[label2]{organization={Department of Radiology, University Hospital Ludwig-Maximilians-University},
            addressline={Marchioninistr. 15},
            city={Munich},
            postcode={81337},
            state={Bavaria},
            country={Germany}}

\begin{abstract}
  Public chest X-ray (CXR) data sets are commonly compressed to a lower bit-depth to reduce their size, potentially hiding subtle diagnostic features.
  In contrast, radiologists apply a windowing operation to the uncompressed image to enhance such subtle features.
  While it was shown that windowing improves classification performance on computed tomography (CT) images, the impact of such an operation on CXR classification performance remains unclear.
  In this study, we show that windowing strongly improves CXR classification performance of machine learning models, and propose WindowNet, a model that learns multiple optimal window settings.
  Our model achieved an average AUC score of 0.812 compared to 0.759 by a commonly used architecture without windowing capabilities on the MIMIC data set.
  Our code is publicly available at \url{https://gitlab.lrz.de/IP/windownet}.
\end{abstract}



\begin{keyword} 
windowing \sep chest X-ray \sep chest radiograph \sep bit-depth \sep classification \sep deep learning



\end{keyword}

\end{frontmatter}


\section{Introduction}
\begin{figure}[h]
  \centering
  \includegraphics[width=\textwidth]{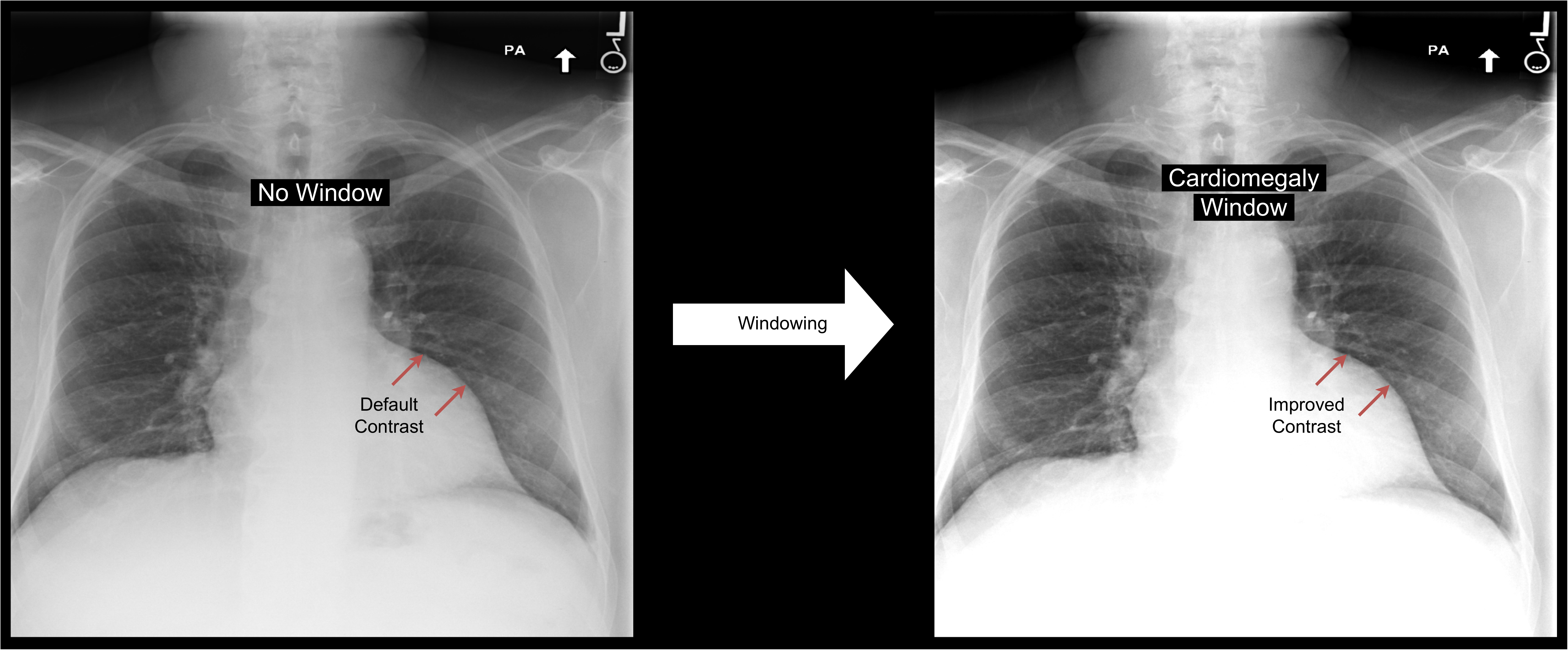}
  \caption{Applying a windowing operation enhances the contrast of particular structures of an image.
    For example, the depicted windowing operation improved cardiomegaly classification performance on the MIMIC data set.}
  \label{fig:window:motivation}
\end{figure}
To better differentiate subtle pathologies, chest X-rays (CXR) are commonly acquired with a high bit-depth.
For example, the images in the MIMIC data set provide 12-bit gray values, see~\cite{johnsonMIMICCXRDeidentifiedPublicly2019}.
However, to reduce the file size and save bandwidth, these images are often compressed to a lower bit-depth.
The Chest X-ray 14 data set, for example, was reduced to 8-bit depth before publication~\citep{wangChestXray8HospitalscaleChest2017}.

Under optimal conditions, the human eye can differentiate between 700 and 900 shades of gray, or 9- to 10-bit depth~\citep{kimpeIncreasingNumberGray2007}.
Hence, radiologists cannot differentiate all 12-bit gray values when inspecting a chest X-ray.
To better identify subtle contrasts, they apply a windowing operation to the image: they increase the contrast by limiting the range of gray tones (see Figure~\ref{fig:window:motivation}).
These windowing operations can be specified by their center (level) and width.

In contrast to chest radiographs, gray values in computed tomography (CT) images are calibrated to represent a specific Hounsfield unit (HU)~\citep{maierMedicalImagingSystems2018}.
For example, a HU value of -1000 corresponds to air, 0 HU to distilled water at standard pressure and temperature, bones range from 400 HU to 3000 HU~\citep{maierMedicalImagingSystems2018}.
To highlight the lung in a chest CT image, one could apply a window with a level of -600 HU and width of 1500 HU~\citep{kazerooniCardiopulmonaryImaging2004}.
In other words, everything below -1350 HU is displayed as black and above 150 HU as white.
Consequently, more distinct gray tone values can be used for the specified range, resulting in a higher contrast.

For CT images, several studies showed that windowing improves classification performance of deep neural networks~\citep{karkiCTWindowTrainable2020,huoStochasticTissueWindow2019,leePracticalWindowSetting2018,kwonTrainableMulticontrastWindowing2020}.
For CXR, no quantitative scale like the Hounsfield Unit exists.
Nevertheless, radiologists window CXR for enhanced contrasts during inspection.
Furthermore, depending on the region of interest, they use different window settings.
This observation leads to the following research questions: does windowing affect chest X-ray classification performance and if yes, can windowing improve it?
To the best of our knowledge, so far, chest X-rays are commonly processed by a deep learning model without applying any windowing operation (for example, \citep{rajpurkarChexnetRadiologistlevelPneumonia2017,wollekAttentionbasedSaliencyMaps2023}).
This study investigates the effect of windowing on chest X-ray classification and proposes a model, WindowNet, that learns optimal windowing settings.

Our contributions are:
\begin{itemize}
  \item We show that a higher bit-depth (8-bit vs. 12-bit) improves chest X-ray classification performance.
  \item We demonstrate that applying a window to the chest radiograph as a pre-processing step increases classification performance.
  \item We propose WindowNet, a chest X-ray classification model that learns optimal windowing settings.
\end{itemize}

\section{Materials and Methods}
\begin{figure}[h]
  \centering
  \includegraphics[width=\textwidth]{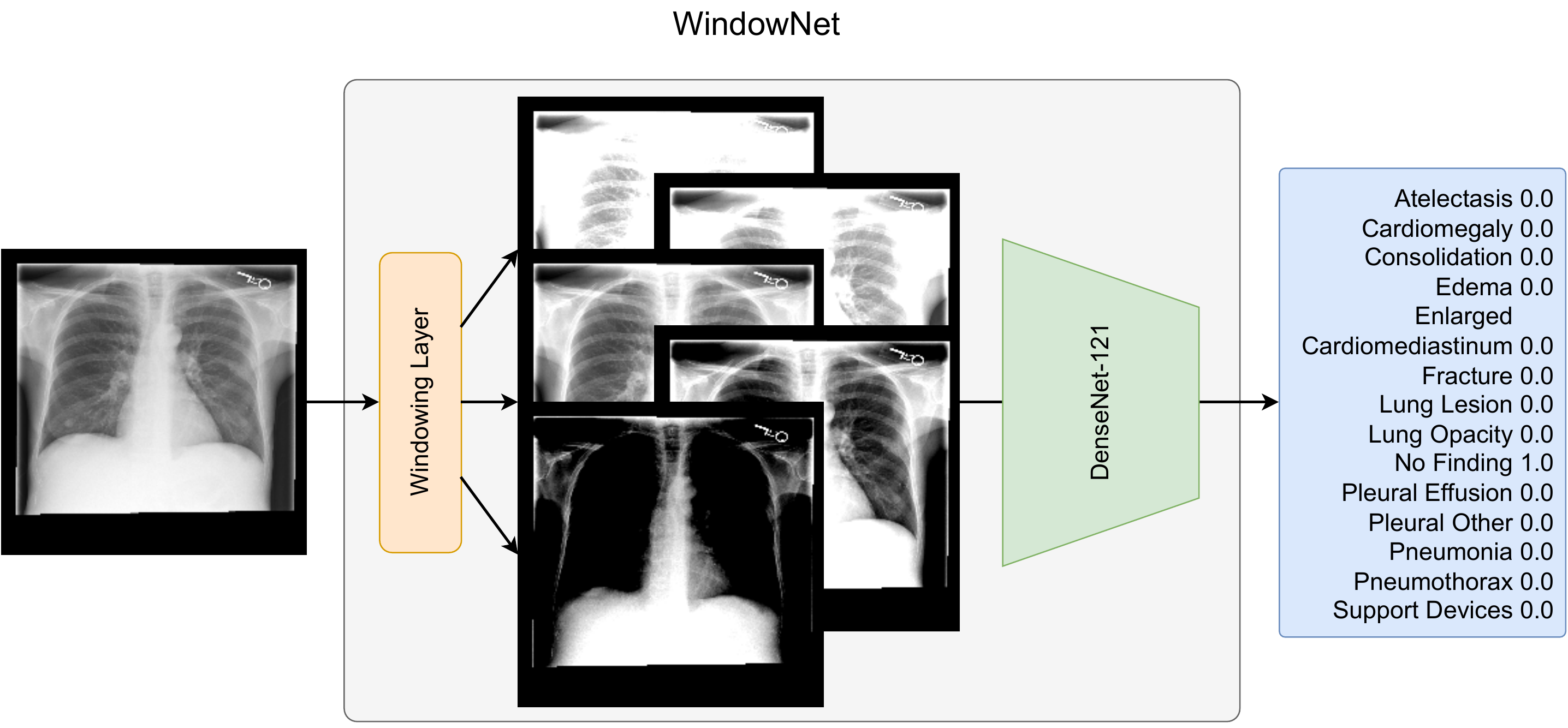}
  \caption{Optimal multi-window chest X-ray classification. Our proposed WindowNet architecture learns to optimize multiple windows for improved classification.}
  \label{fig:window:architecture}
\end{figure}

\subsection{Data Set}
To investigate the importance of windowing on chest X-ray classification we selected the MIMIC data set, as it is the only publicly available, large-scale chest X-ray data set with full bit-depth~\citep{johnsonMIMICCXRDeidentifiedPublicly2019}.
The MIMIC data set provides chest radiographs in the original Digital Imaging and Communications in Medicine (DICOM) format with 12-bit depth gray values, containing 377,110 frontal and lateral images from 65,379 patients.
The images have been labeled according to the 14 CheXpert classes: atelectasis, cardiomegaly, consolidation, edema, enlarged cardiomediastinum, fracture, lung lesion, lung opacity, no finding, pleural effusion, pleural other, pneumonia, pneumothorax, and support devices~\citep{irvinChexpertLargeChest2019}.
In our experiments, we used the provided training, validation, and test splits.
During pre-processing, the images were resized to $224\times224$ pixels.

\subsection{Architectures}
\subsubsection{Baseline}
As a baseline model (Baseline) for all experiments, we used a DenseNet-121~\citep{huangDenselyConnectedConvolutional2017} pre-trained on ImageNet~\citep{ImageNet} that is commonly used for chest X-ray classification~\citep{rajpurkarChexnetRadiologistlevelPneumonia2017,wollekKneeCannotHave2023submitted,xiaoDelvingMaskedAutoencoders2023}.
For fine-tuning, we replaced the classification layer with a 14-dimensional fully-connected layer.

\subsubsection{WindowNet}
To incorporate windowing into the model architecture, we extended the baseline architecture by prepending a windowing layer, as illustrated in Figure~\ref{fig:window:architecture}.
In the following, we refer to this model as WindowNet.

We implemented the windowing operation as a $1\times1$ convolution with clamping, similar to~\citep{leePracticalWindowSetting2018}.
This implementation of windowing utilizing convolutional kernels enables the model to learn and use multiple windows in parallel.
As the pre-trained DenseNet-121 expects three input channels, we added an additional $1\times1$ convolution with three output channels after the windowing operation.
Following the windowing layer, the images are scaled to the floating point range $(0.0, 255.0)$ and then normalized according to the ImageNet mean and standard deviation.

\subsubsection{Training}
Both models were trained with binary cross-entropy loss, AdamW optimization with a learning rate of 1e-4~\citep{loshchilovDecoupledWeightDecay2019}, and a batch size of 32.
During training, the learning rate was divided by a factor of 10 if the validation loss did not improve in three consecutive epochs.
The training was stopped if the validation loss did not improve after 5 consecutive epochs.
The final models were selected based on the checkpoint with the highest mean validation area under the receiver operating characteristic curve (AUC).

\subsection{Experiments}
\subsubsection{8-Bit vs. 12-Bit}
As applying a windowing operation in our experiments required a higher initial bit-depth than conventionally used for chest X-ray image classification, we first tested the effect of bit-depth on classification performance.
We trained the baseline model with 8-bit and 12-bit depth and compared mean, and class-wise AUC scores.
In both settings no windowing operation was applied.
However, the 12-bit images were still scaled to the floating point range $(0.0, 255.0)$.
In both settings, the images were normalized according to the ImageNet mean and standard deviation.

\subsubsection{Single Fixed Window}
To investigate whether windowing has an effect on classification performance, we trained the baseline model with a single fixed windowing operation applied to the 12-bit CXRs.
After windowing, the images were scaled to have a maximum value of 255 and normalized according to the ImageNet mean and standard deviation.

For windowing, we use a fixed window level of 100, and levels ranging from 250 to 3500 in steps of 250.
All levels were combined with fixed window widths of 500, 1000, 1500, 2000, and 3000.
For evaluation, we compared the mean and class-wise AUCs of each model to the baseline with no windowing, i.e., a window level of 2048 and width of 4096.

\subsubsection{Trainable Multi-Windowing}
To test if end-to-end optimized windows improve chest X-ray classification performance we compared our proposed WindowNet to the baseline and a modified WindowNet without clamping in the windowing layer (No Windowing), i.e., a conventional $1\times1$ convolutional layer.

In our experiments, we used 14 windows based on the set of class-wise top-3 windows found during the single window experiment and the additional full-range ``window''.
The selection was based on the validation results.
We initialized the learnable windows with the resulting windows (level, width): (100, 3000), (1250, 1000), (1500, 3000), (1750, 2000), (1750, 3000), (2000, 2000), (2250, 2000), (2250, 3000), (2500, 2000), (2500, 3000), (2750, 3000), (3250, 1000), (750, 3000), and (2048, 4096).
The comparison model, ``No Windowing'', having a conventional $1\times1$ convolution, was default initialized using kaiming initalization~\citep{heDelvingDeepRectifiers2015}.

\section{Theory}
A windowing operation can be described by its center (window level) and width (window width).
Formally, the windowing operation applied to a pixel value $px$ can be defined as:
\begin{align*}
  \mathit{window}(px) &= \min(\max(px, L), U), \\
  U &= WL + \frac{WW}{2}, \\
  L &= WL - \frac{WW}{2}. \\
\end{align*}
Where \textit{U} is the upper limit and \textit{L} the lower limit of the window defined by the window level \textit{WL} and window width \textit{WW}.

For efficient training, the windowing operation can be re-written using a clamped $1\times1$ convolution.
Here, the weight matrix is initialized as $W = \frac{U}{WW}$ and the bias term as $b = -\frac{U}{WW}L$, similar to~\citep{leePracticalWindowSetting2018}:
\begin{align*}
  \min\left(\max\left(Wx + b, 0\right), U\right) &= \min\left(\max\left(\frac{U}{WW}x - \frac{U}{WW}L, 0\right), U\right) \\
                                                 &= \min\left(\max\left(\frac{U}{WW}\left(x - L \right), 0\right), U\right) \\
                                                &= \min\left(\max(x - L, 0), U\right) \\
                                                 &= \min(\max(x, L), U). \\
\end{align*}
To recover the window level and width after training, we compute:
\begin{align*}
  WW &= \frac{U}{W}, \\
  WL &= -\frac{b}{W} + \frac{WW}{2}. \\
\end{align*}

\section{Results}
\subsection{8-Bit vs. 12-Bit}
\begin{table}
  \centering
\begin{tabular}{lll}
{Finding}                  &  8-Bit & 12-Bit \\
\midrule
Atelectasis                &  \textbf{0.751} &  0.749 \\
Cardiomegaly               &  0.770 &  \textbf{0.774} \\
Consolidation              &  0.740 &  \textbf{0.742} \\
Edema                      &  0.831 &  \textbf{0.833} \\
Enlarged Cardiomediastinum &  0.691 &  \textbf{0.701} \\
Fracture                   &  0.664 &  \textbf{0.710} \\
Lung Lesion                &  0.680 &  \textbf{0.682} \\
Lung Opacity               &  0.680 &  \textbf{0.690} \\
No Finding                 &  0.789 &  \textbf{0.797} \\
Pleural Effusion           &  \textbf{0.883} &  0.879 \\
Pleural Other              &  0.823 &  \textbf{0.831} \\
Pneumonia                  &  0.659 &  \textbf{0.698} \\
Pneumothorax               &  0.802 &  \textbf{0.828} \\
Support Devices            &  0.868 &  \textbf{0.888} \\
\midrule
Mean                       &  0.759 &  \textbf{0.772} \\
\end{tabular}

\caption{Effect of bit-depth on chest X-ray classification performance.
  A higher bit-depth improved AUC values for most (12/14) classes.
  Higher values are highlighted in bold.
  }
  \label{tab:8bit}
\end{table}

The classification AUCs, when trained with 8-bit or 12-bit depth, are shown in Table~\ref{tab:8bit}.
Training with 12-bit images improved the average classification performance compared to 8-bit images (0.772 vs. 0.759 AUC).
Also, most (12/14) class-wise AUCs increased when training with a higher bit-depth.
The only exceptions where atelectasis and pleural effusion, where training with 8-bit images resulted in slightly higher AUC with 0.751 vs. 0.749 and 0.883 vs. 0.879, respectively.

\subsection{Single Fixed Window}
\begin{table}
  \centering
  \begin{tabular}{lll}
Finding  &	No Window &	Best Fixed Window \\
\midrule
Atelectasis                &  0.749 (2048, 4096) &  \textbf{0.757} (2750, 3000) \\
Cardiomegaly               &  0.774 (2048, 4096) &  \textbf{0.786} (1750, 3000) \\
Consolidation              &  0.742 (2048, 4096) &  \textbf{0.744} (2500, 3000) \\
Edema                      &  0.833 (2048, 4096) &  \textbf{0.841} (1750, 3000) \\
Enlarged Cardiom.          &  0.701 (2048, 4096) &  \textbf{0.734} (2250, 3000) \\
Fracture                   &  \textbf{0.710} (2048, 4096) &  0.706 (1000, 3000) \\
Lung Lesion                &  0.682 (2048, 4096) &  \textbf{0.720} (2500, 3000) \\
Lung Opacity               &  \textbf{0.690} (2048, 4096) &  \textbf{0.690} (2250, 3000) \\
No Finding                 &  0.797 (2048, 4096) &  \textbf{0.804} (2500, 3000) \\
Pleural Effusion           &  0.879 (2048, 4096) &  \textbf{0.888} (2500, 3000) \\
Pleural Other              &  0.831 (2048, 4096) &  \textbf{0.850} (2750, 3000) \\
Pneumonia                  &  \textbf{0.698} (2048, 4096) &  0.690 (1750, 3000) \\
Pneumothorax               &  0.828 (2048, 4096) &  \textbf{0.832} (1750, 3000) \\
Support Devices            &  0.888 (2048, 4096) &  \textbf{0.889} (2750, 3000) \\
\midrule
Mean                       &  0.772 (2048, 4096) &  \textbf{0.775} (2500, 3000) \\
\end{tabular}

  \caption{Effect of fixed windowing on chest X-ray classification AUCs.
    For each finding, the best performing window and the baseline without no windowing are reported.
    Higher AUCs values are highlighted in bold.
    Enlarged Cardiom. = enlarged cardiomediastinum.
  }\label{tab:gridwindow}
\end{table}

The results of training with fixed window chest X-rays are reported in Table~\ref{tab:gridwindow}.
They demonstrate that windowing improved chest X-ray classification AUCs for most classes (12/14) except for fracture and pneumonia with AUCs of 0.710 vs. 0.706 and 0.698 vs. 0.690, respectively.
On average, the window with level 2500 and width 3000 performed slightly better than the full range with an AUC of 0.775 vs. 0.772.
Across all windows, a window width of 3000 performed best with varying window levels.
\begin{table}
  \centering
\begin{tabular}{lrrrrr}
Window                     & None (Baseline) &	\#1 &	\#2 &	\#3 &	\#4 \\
\midrule
Level                      &   2048 &   2500 &   1750 &   2750 &   2250 \\
Width                      &   4096 &   3000 &   3000 &   3000 &   3000 \\
                           &                   &                   &                   &                   & \\
Finding                    &                   &                   &                   &                   & \\
\midrule
Atelectasis                &  0.749 &  0.756 &  0.753 &  0.749 &  \textbf{0.757} \\
Cardiomegaly               &  0.774 &  0.783 &  \textbf{0.786} &  0.774 &  0.777 \\
Consolidation              &  0.742 &  \textbf{0.744} &  0.743 &  0.742 &  0.740 \\
Edema                      &  0.833 &  0.830 &  \textbf{0.841} &  0.833 &  0.831 \\
Enlarged Cardiom.          &  0.701 &  \textbf{0.710} &  0.700 &  0.701 &  0.686 \\
Fracture                   &  \textbf{0.710} &  0.695 &  0.670 &  \textbf{0.710} &  0.669 \\
Lung Lesion                &  0.682 &  \textbf{0.720} &  0.710 &  0.682 &  0.700 \\
Lung Opacity               &  \textbf{0.690} &  0.683 &  0.686 &  \textbf{0.690} &  0.684 \\
No Finding                 &  0.797 &  \textbf{0.804} &  0.800 &  0.797 &  0.798 \\
Pleural Effusion           &  0.879 &  \textbf{0.888} &  0.883 &  0.879 &  0.885 \\
Pleural Other              &  0.831 &  0.841 &  0.820 &  0.831 &  \textbf{0.850} \\
Pneumonia                  &  \textbf{0.698} &  0.686 &  0.690 &  \textbf{0.698} &  0.683 \\
Pneumothorax               &  0.828 &  0.822 &  \textbf{0.832} &  0.828 &  0.809 \\
Support Devices            &  0.888 &  0.887 &  0.887 &  0.888 &  \textbf{0.889} \\
\midrule
Mean (Validation)          &  0.804 &  \textbf{0.807} &  0.802 &  0.805 &  0.803 \\
Mean (Test)                &  0.772 &  \textbf{0.775} &  0.772 &  0.772 &  0.768 \\
\end{tabular}

  \caption{Best fixed single window settings for chest X-ray classification found during grid search.
    The class-wise AUCs of the four best performing windows (Window 1-4) and the baseline without windowing are reported.
    Additionally, mean validation AUCs are provided.
    Highest AUC values are highlighted in bold.
    Enlarged Cardiom. = enlarged cardiomediastinum.}
  \label{tab:gridbestwindow}
\end{table}

A comparison of the four best-performing windows to the baseline is shown in Table~\ref{tab:gridbestwindow}.
All five settings achieved similar average AUC scores.
No single window performed consistently better across all classes, suggesting that multiple windows could overall improve the classification performance.

\subsection{Trainable Multi-Windowing}
\begin{figure}
  \centering
  \includegraphics[width=\textwidth]{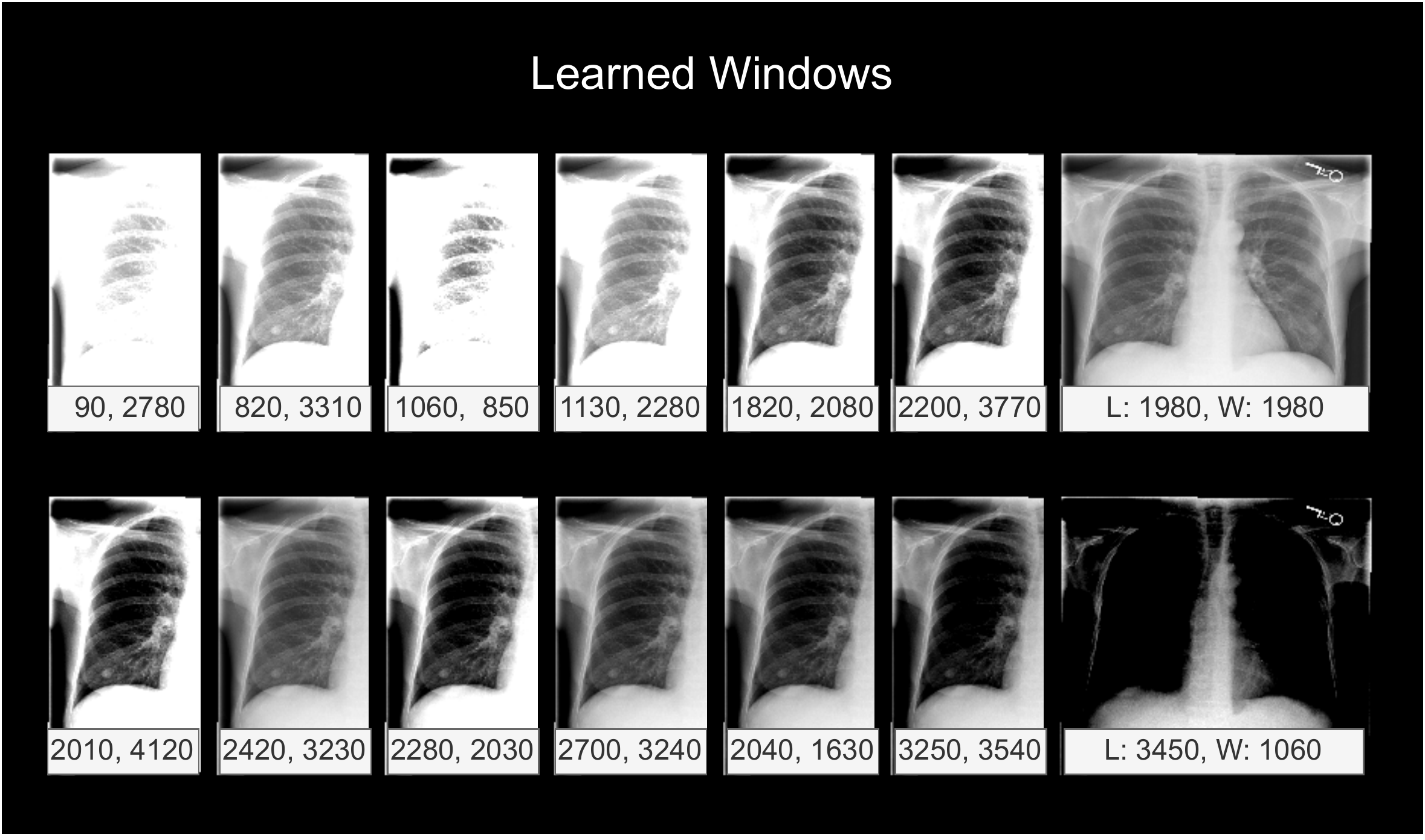}
  \caption{Windows learned during the training of WindowNet. For window initialization the following window levels (L) and widths (W) were used (level, width): (100, 3000), (1250, 1000), (1500, 3000), (1750, 2000), (1750, 3000), (2000, 2000), (2250, 2000), (2250, 3000), (2500, 2000), (2500, 3000), (2750, 3000), (3250, 1000), (750, 3000), and no window (2048, 4096).}
  \label{fig:window:learned_windows}
\end{figure}
\begin{table}
  \centering
  \begin{tabular}{lrrr}
{Finding}                  &   8-Bit &   No Windowing (12-Bit) &  WindowNet (12-Bit)    \\
\midrule
Atelectasis                &   0.751 &   0.812 &             \textbf{0.829}\\
Cardiomegaly               &   0.770 &   0.814 &             \textbf{0.827}\\
Consolidation              &   0.740 &   0.808 &             \textbf{0.823}\\
Edema                      &   0.831 &   0.891 &             \textbf{0.897} \\
Enlarged Cardiom.          &   0.691 &   0.745 &             \textbf{0.764}\\
Fracture                   &   \textbf{0.664} &   0.619 &             0.615 \\
Lung Lesion                &   0.680 &   0.701 &             \textbf{0.744} \\
Lung Opacity               &   0.680 &   0.726 &             \textbf{0.745}\\
No Finding                 &   0.789 &   0.855 &             \textbf{0.859} \\
Pleural Effusion           &   0.883 &   0.909 &             \textbf{0.918}\\
Pleural Other              &   \textbf{0.823} &   0.721 &             0.793\\
Pneumonia                  &   0.659 &   0.731 &             \textbf{0.750}\\
Pneumothorax               &   0.802 &   0.830 &             \textbf{0.886} \\
Support Devices            &   0.868 &   0.897 &             \textbf{0.918}\\
\midrule
Mean                       &   0.759 &   0.790 &             \textbf{0.812} \\
  \end{tabular}
  \caption{Comparison of baseline (8-Bit), WindowNet without windowing (``No Windowing 12-Bit'') and WindowNet (12-Bit) AUCs for chest X-ray classification.
    Higher values are highlighted in bold.
    Enlarged Cardiom. = enlarged cardiomediastinum.}
  \label{tab:learnedwindow}
\end{table}

The effect of learning multiple optimal windows using our proposed WindowNet is reported in Table~\ref{tab:learnedwindow}, comparing it to the baseline and the WindowNet architecture without windowing (``No Windowing'').
Overall, WindowNet performed considerably better with an average AUC of 0.812 compared to 0.750 of the 8-bit baseline.
When compared to a conventional $1\times1$ convolution in the WindowNet architecture (``No Windowing'') the results demonstrate the improvement of windowing with an average AUC of 0.812 vs. 0.790.

For nearly all classes (12/14) our proposed WindowNet model achieved a higher AUC.
For example, pneumothorax classification AUC improved from 0.802 to 0.886 with windowing.
Only for the fracture and pleural other class the baseline model performed better with an AUC of 0.664 vs. 0.615 and 0.823 vs. 0.793, respectively.

The windows learned after training are shown in Figure~\ref{fig:window:learned_windows}.
The model learned a diverse set of windows, with levels from 90 to 3450 and widths from 850 to 4120.



\section{Discussion}
In this study, we investigated the importance of windowing, inspired by radiologists.
Our results show that our proposed multi-windowing model, WindowNet, considerably outperformed a popular baseline architecture with a mean AUC of 0.812 compared to 0.759 (see Table~\ref{tab:learnedwindow}).
As a necessary pre-condition, we also demonstrated that the common bit-depth reduction negatively affected classification performance (0.759 vs. 0.772 AUC), as seen in Table~\ref{tab:8bit}.

Similarly to related work in the CT domain~\citep{leePracticalWindowSetting2018, karkiCTWindowTrainable2020, kwonTrainableMulticontrastWindowing2020}, our results show that windowing is a useful pre-processing step for neural networks operating on chest X-rays.
These findings are also in line with the observed manual windowing performed by radiologists in their daily practice.
In addition, like radiologists apply multiple windows when inspecting a single image, no single window was better across classes, including not windowing at all (see Table~\ref{tab:gridwindow}).

When comparing our proposed WindowNet with the same architecture but without windowing, in other words, a conventional $1\times1$ convolution, our results showed that the windowing operation is an important aspect of the architecture (see Table~\ref{tab:learnedwindow}).
When inspecting the learned windows, see Figure~\ref{fig:window:learned_windows}, the windows converged to 14 different settings.
This provides further evidence that multiple windows are important for classification performance.

While our study's results are promising, limitations include the exploratory nature of the study and the evaluation on a data set from a single institution, due to lack of other high bit depth public data sets.
Further research is needed to show generalization to other data sets and institutions.
Another limitation is that the model learns general windowing settings.
In contrast, radiologists adapt the windowing setting based on the specific image.
Future work could investigate an image-based window setting prediction layer.

In conclusion, we believe our work offers an important contribution to the field of computer vision and radiology by demonstrating that multi-windowing strongly improves chest X-ray classification performance, as shown by our proposed model, WindowNet.

\section{Acknowledgments}
This work was supported in part by the German federal ministry of health’s program for digital innovations for the improvement of patient-centered care in healthcare [grant agreement no. 2520DAT920].



\bibliographystyle{elsarticle-harv}

\end{document}